\documentclass[showpacs,preprintnumbers]{revtex4}

\usepackage{mathrsfs}
\usepackage[latin1]{inputenc}
\usepackage[gen]{eurosym}
\usepackage{graphicx}
\usepackage{floatflt}
\usepackage{amsmath}
\usepackage{bbm}

\usepackage{color}

\usepackage{relsize}

\usepackage{fancybox}

\DeclareMathAlphabet{\mathpzc}{OT1}{pzc}{m}{it}

\usepackage{stmaryrd}

\usepackage{amssymb}

\usepackage{amsbsy}

\usepackage{amssymb}
\usepackage{cancel}

\usepackage{subfigure}

\usepackage[abs]{overpic}

\usepackage{amssymb}

\usepackage{epigraph}
\begin{document}

\title{$U(2)_A \times U(2)_V$-symmetric fixed point from the Functional Renormalization Group}
\author{Mara Grahl }
\affiliation{Institute for Theoretical Physics, Goethe University,
Max-von-Laue-Str.\ 1, D--60438 Frankfurt am Main, Germany }

\begin{abstract}

The existence of an $U(2)_A \times U(2)_V$-symmetric fixed point in the
chiral linear sigma model is confirmed using the Functional Renormalization Group (FRG).
Its stability properties and the implications for the order of the 
chiral phase transition of two-flavor quantum chromodynamics (QCD) are discussed.
Furthermore, several technical conclusions are drawn from the comparison 
with the results of resummed loop expansions.

\end{abstract}
\maketitle

\section{Introduction}

Apart from other methods, our current understanding of QCD in the 
nonperturbative regime is strongly based 
on lattice gauge theory and 
effective models \cite{Leupold:2011zz}.
These complementary approaches
are compared to each other
for reasons
of crosscheck and systematic improvement \cite{Pawlowski:2010ht,Herbst:2013ufa}. 
Despite all efforts the order of the chiral phase 
transition of QCD with two
massless flavors has not been rigorously determined yet, and
the interest in a reliable prediction 
remains strong.
The case of two massless (or light, respectively) flavors at vanishing baryonic chemical potential
is of particular interest for lattice studies due to  
the comprehensive predictions of effective models \cite{Cossu:2013uua,Bhattacharya:2014ara}.
The possible existence of a second-order chiral phase transition, as well
as the corresponding universality class, can be investigated from 
the effective theory for the chiral 
condensate \cite{Pisarski:1983ms,Berges:1997eu,Berges:1998sd,Butti:2003nu,Calabrese:2004uk,Braun:2010vd,Fukushima:2010ji,Grahl:2013pba,Pelissetto:2013hqa,Aoki:2013zfa,Meggiolaro:2013swa,Nakayama:2014sba,Fejos:2014qga}.
Using the $[\bar{2},2]+[2,\bar{2}]$
representation of $SU(2)_L \times SU(2)_R$ \cite{Paterson:1980fc},
we can take into account the scalar mesons ($\sigma$ and $\vec{a}_0$) 
as well as the pseudoscalar mesons ($\eta$ and $\vec{\pi}$) by 
writing down the most general Lagrangian invariant under chiral symmetry.
For the full symmetry, $U(2)_A \times U(2)_V \simeq U(1)_A \times U(1)_V \times 
\left[ SU(2)/Z(2) \right]_L \times \left[ SU(2)/Z(2) \right]_R$,
taking into account all linearly independent invariants up to eighth 
polynomial order in the fields,
this Lagrangian can be written as \cite{Pisarski:1983ms,Berges:1997eu,GrahlDiss,Patkos:2012ex}
\begin{gather}
 \mathscr{L} = \frac{Z}{2} {\rm Tr} (\partial_{\mu} \Phi^{\dagger} ) 
(\partial_{\mu} \Phi) +r {\rm Tr} \Phi^{\dagger} \Phi 
+  g_1 ({\rm Tr} \Phi^{\dagger} \Phi)^2 + g_2 \xi + g_3 ({\rm Tr} \Phi^{\dagger} \Phi)^3 \nonumber\\
+g_4 ({\rm Tr} \Phi^{\dagger} \Phi) \xi+  g_5 {\rm Tr} ( \Phi^{\dagger} \Phi)^4 
+g_6 ({\rm Tr} \Phi^{\dagger} \Phi)^4 +g_7 ({\rm Tr} \Phi^{\dagger} \Phi)^2 {\rm Tr} ( \Phi^{\dagger} \Phi)^2    \; , \label{lsp}
\end{gather}
where 
$ \Phi = \left(\sigma + i \eta \right) t_0 + \vec{t} \cdot \left(\vec{a}_0
  + i \vec{\pi} \right)$, with $t_a$ denoting the generators of $U(2)$ normalized such that ${\rm Tr}(t_a t_b) \equiv 1$ \cite{Grahl:2013pba}. 
Furthermore, 
\begin{gather*}
{\rm Tr} \Phi^{\dagger} \Phi = \sum_{i} \phi_i^2 \equiv 2 \rho\;, \ \phi_i \equiv \sigma,\vec{\pi},\eta,\vec{a}_0 \;, \\
\frac{1}{2}{\rm Tr} ( \Phi^{\dagger} \Phi)^2 - \rho^2 =  \left(\sigma^2 +\vec{\pi}^2 \right) \left(\eta^2 +
     \vec{a}_0^2 \right) - \left(\sigma \eta - \vec{\pi} \cdot \vec{a}_0 \right)^2   \equiv  \xi\;.
\end{gather*}
\noindent
We omit derivate couplings since we will only discuss the local-potential
approximation (LPA, $Z=1$) and, respectively, its minimal extension allowing for a
field-independent wave-function renormalization factor $Z$ (LPA').
We note that the invariants $({\rm Tr} \Phi^{\dagger} \Phi) {\rm Tr} ( \Phi^{\dagger} \Phi)^3$, 
$\left({\rm Tr} ( \Phi^{\dagger} \Phi)^2 \right)^2$, and ${\rm Tr} ( \Phi^{\dagger} \Phi)^3$
do not yield further linearly independent contributions 
to Eq.\ (\ref{lsp}). \\
In this paper we focus on the case where the axial $U(1)_A$ symmetry, which is anomalously broken
at vanishing temperature, has already been restored at the critical temperature $T_c$. 
Therefore we do not take account of $U(1)_A$-breaking terms in Eq.\ (\ref{lsp}).
For studies concerning the opposite scenario in which the anomaly remains present at $T_c$ we
refer to Refs.\ \cite{Grahl:2013pba,Pawlowski:1996ch,Fischer:2011pk,Pelissetto:2013hqa,Schaefer:2013isa,Mitter:2013fxa,Aoki:2013zfa,Meggiolaro:2013swa}. 
The long-standing question which of the both scenarios is actually realized
is subject to an ongoing debate. The latest lattice results are quite controversial: whereas 
the case of restored anomaly is advocated by Refs.\ \cite{Aoki:2012yj,Cossu:2013uua}, the opposite
scenario is favored by Refs.\ \cite{Sharma:2013nva,Bhattacharya:2014ara}.
The predictions of effective theories for the chiral condensate are summarized in the following. \\ 
The existence of an infrared-stable (IR-stable) fixed point in the RG flow
of the effective theory for the order parameter is a necessary 
condition for a second-order phase transition to occur.
If this scenario is realized or not depends on the initial values
for the parameters in the ultraviolet (UV) limit determined by
the underlying microscopic theory. Therefore, the RG analysis
serves to either rule out the existence of a second-order phase transition
or to confirm its possible existence. \\
If the anomaly strength exceeds the cut-off scale, a phase transition 
of second order in the $O(4)$ universality class is 
predicted \cite{Pisarski:1983ms,Jungnickel:1995fp,Grahl:2013pba}. 
The case of small anomaly strength is subtle. The anomaly yields two independent quadratic
mass terms. In Landau theory, i.e., at mean-field
level, it is evident that such a situation corresponds to a multicritical point with at least two relevant
scaling variables. This is used as an argument in Ref.\ \cite{Pelissetto:2013hqa} to rule out 
a second-order phase transition with temperature being the only relevant scaling variable. 
However, in consistence with Refs.\ \cite{Grahl:2013pba,GrahlDiss}, we argue that the inclusion of 
fluctuations can, in principle, lead to a IR-stable fixed point corresponding to exactly such a scenario.
Although associated with unphysical masses in the approximation considered, there in fact exists
an (unphysical) $SU(2)_A \times U(2)_V$-symmetric, IR-stable fixed point
exemplifying our consideration. 
This observation extends the critical reinvestigation of the
standard criterions used for ruling out continuous transitions presented in Ref.\ \cite{Przy}. The 
latter particularly points out that the irreducibility of a representation is not strictly ruling
out a second-order phase transition associated with a single relevant scaling variable.
In the absence of the anomaly there is strong evidence 
from Refs.\ \cite{Pelissetto:2013hqa,Nakayama:2014sba} for the 
existence of a second-order phase transition belonging to the
$U(2)_V \times U(2)_A$ universality class. The existence and properties of the corresponding
fixed point will be discussed in the remainder of this paper. \\
Ref.\ \cite{Pelissetto:2013hqa} uses a resummed loop expansion at fixed spatial dimension, $D=3$, based on
the $\overline{MS}$ and the \textit{MZM} scheme, respectively. The discovered IR-stable, $U(2)_V \times U(2)_A$-symmetric
fixed point corresponds to an anomalous dimension of $\eta \sim 0.12$. Previous studies in the 
framework of the $\epsilon$-expansion ($\epsilon = 4-D$) failed to find the fixed point \cite{Pisarski:1983ms,Calabrese:2004uk}.
It is an important question why this is the case. 
A plausible explanation is given in Ref.\ \cite{Pelissetto:2013hqa}: the 
fixed point only exists near $D=3$. One might wonder, however, if the
resummation scheme and the loop-order also play a role.
With our FRG investigation presented in Secs.\ \ref{frgres}--\ref{staban} we demonstrate that 
the existence not only depends on the fixed spatial dimension, but also on the
way how nonperturbative corrections are included. \\
Due to the converging correlation length at a second-order phase transition we can 
work in the dimensionally reduced theory \cite{DimRed2ndorder}.

\section{Fixed points from FRG}
\label{frgres}

Assuming an homogeneous condensate, and using the Litim regulator,
the Wetterich equation for the potential of the truncation (\ref{lsp})
is given by
\begin{gather}
 \frac{\partial U_k}{\partial k} = \frac{2\pi^{D/2} k^{D+1} Z_k}{D \ \Gamma(D/2) (2\pi)^D}  \left(1- \frac{\eta}{2+D}\right) \sum_{i} 
 \frac{1}{Z_k k^2 + M_i^2} \; ,  \label{fleq}
\end{gather}
\noindent
where $\mathscr{L}_{k} = \frac{1}{2} Z_k Tr (\partial_{\mu} \Phi^{\dagger}) (\partial_{\mu} \Phi)+U_k$,  
with $\mathscr{L}_{k=\Lambda}=\mathscr{L}$ defining the bare Lagrangian in the UV limit.
$M_i^2$ denote the eigenvalues of the mass matrix
\begin{gather}
 M_{ij} \equiv \frac{\partial^2 U_k}{\partial \phi_i \partial \phi_j}
 \; , \; \; i,j=1,\ldots,8 \; . \label{mm}
\end{gather}
\noindent
The anomalous dimension, $\eta$, is determined from the relation 
\begin{gather}
\eta_k = - Z_k^{-1} k \frac{\partial Z_k}{\partial k} \; , \; \lim_{k \to 0} \eta_k = \eta\;.
\end{gather}
The flow equation for $Z_k$ is derived from the second derivative of the effective action with
respect to the fields and evaluated at the global minimum of the potential \cite{Kopietz:FRG}. 
For our purposes we can restrict our discussion of the LPA' to the truncation 
$U_k (\rho, \xi) \equiv V(\rho) + W(\rho) \xi$, which is suited up to sextic
truncation order ($g_5 = g_6 = g_7 = 0$).
Setting $D=3$, in agreement with Ref.\ \cite{Berges:2000ew} we obtain
\begin{gather}
\eta_k = \frac{2}{3 \pi^2   [1+\bar{V}_k^{'}(\bar{\rho}_{0,k})]^2  } \left(  \frac{4 \bar{\rho}_{0,k} 
\bar{W}_k(\bar{\rho}_{0,k})^2}{ [1+ 4 \bar{W}_k(\bar{\rho}_{0,k}) \bar{\rho}_{0,k} + \bar{V}_k^{'}(\bar{\rho}_{0,k})]^2 }
        + \frac{ \bar{\rho}_{0,k} \bar{V}_k^{''}(\bar{\rho}_{0,k})^2 }{
        [1+ \bar{V}^{'}(\bar{\rho}_{0,k}) +2\bar{\rho}_{0,k} \bar{V}_k^{''}(\bar{\rho}_{0,k})]^2 } \right) \; ,
\end{gather}
\noindent
where we introduced rescaled variables (labeled by a bar),
\begin{gather*}
\bar{U} = k^{-D} U \; , \; \bar{\rho}= Z  k^{2-D} \rho \; , \; \bar{\xi} = Z^2 k^{4-2D} \xi
\; , \; \bar{V} = k^{-D} V \; , \; \bar{W} = Z^{-2} k^{D-4} W \; ,
\end{gather*}
\noindent
and denoted the global minimum of $U_k$ by $\rho_0$ (assuming $\xi_0 = 0$). \\ 
The flow equations for the rescaled parameters of Eq.\ (\ref{lsp}) are 
derived similar to Refs.\ \cite{Fukushima:2010ji,Grahl:2013pba},
not listed explicitly here.
The numerically determined fixed points for sextic truncation order are listed in Table \ref{tabfp6}, those
for octic truncation order in Table \ref{tabfp8}. We proceed with a detailed analysis of their stability
properties and the resultant implications in Sec.\ \ref{staban}.
\begin{table}
\caption{\label{tabfp6}Fixed points in sextic truncation order (for the LPA denoted by $F_i^{(6)}$, for the 
LPA' by $F_i'^{(6)}$). $D=3$.}
\begin{ruledtabular}
\begin{tabular}{l|l l l l l | l   }
 $F$ & $\bar{r}$ & $\bar{g}_1$ & $\bar{g}_2$ & $\bar{g}_3$ & $\bar{g}_4$ & $\eta$ \hspace{1cm}  \\ \hline
 $F_0^{(6)}$, $F_0'^{(6)}$ & 0 & 0 & 0 & 0 & 0 & 0 \\
 $F_1^{(6)}$ & -0.1316 & 0.0827 & 0.8586 & 0.2091 & 0.2161 & 0  \\
 $F_1'^{(6)}$ & -0.1251 & 0.0795 & 0.8447 & 0.1981 & 0.1876 & 0.0334  \\
 $F_2^{(6)}$ & -0.103 & 0.3334 & -0.9411 & 0.307 & -0.7154 &  0 \\
 $F_2'^{(6)}$ & -0.0938 & 0.3151 & -0.8981 & 0.2634 & -0.6024 &  0.0529 \\
 $F_3^{(6)}$ & -0.1355 & 0.2132 & 0 & 0.1285 & 0 & 0 \\
 $F_3'^{(6)}$ & -0.1317 & 0.2103 & 0 & 0.123 & 0 & 0.0195 \\
\end{tabular}
\end{ruledtabular}
\begin{ruledtabular}
\begin{tabular}{l|c | c  }
 $F$ & stability-matrix eigenvalues & nonzero $\bar{M}_i^2$  \\ \hline
 $F_0^{(6)}$, $F_0'^{(6)}$ & \{-2,-1,-1,0,0\} & -- \\
 $F_1^{(6)}$ &  \{15.6603,0.6245+3.5342 i,0.6245-3.5342 i,1.6306,-1.3743\} & \{0.8246,0.6434,0.6434,0.6434\} \\
 $F_1'^{(6)}$ &  \{14.5059,0.5839+3.2722 i,0.5839-3.2722 i,1.5485,-1.3614\} & \{0.7823,0.6261,0.6261,0.6261\}  \\
 $F_2^{(6)}$ & \{13.2219,1.1882+2.1481 i,1.1882-2.1481 i,-1.5108,1.3732\} & \{0.4750,-0.2707,-0.2707,-0.2707\} \\
 $F_2'^{(6)}$ &  \{11.6716,1.0622+1.874 i,1.0622-1.874 i,-1.5279,1.3464\} & \{0.4272,-0.2502,-0.2502,-0.2502\} \\
 $F_3^{(6)}$ & \{12.9247,8.125,1.5092,-1.3798,-0.5034\} & \{0.6442\} \\
 $F_3'^{(6)}$ & \{12.3598,7.7931,1.4673,-1.3745,-0.4802\} & \{0.6233\} \\
\end{tabular}
\end{ruledtabular}
\end{table}

\begin{table}
\caption{\label{tabfp8}Fixed points for the LPA in octic truncation order. $D=3$. }
\begin{ruledtabular}
\begin{tabular}{l|l l l l l l l l  }
 $F$ & $\bar{r}$ & $\bar{g}_1$ & $\bar{g}_2$ & $\bar{g}_3$ & $\bar{g}_4$ & $\bar{g}_5$ & $\bar{g}_6$ & $\bar{g}_7$  \\ \hline
 $F_0^{(8)}$ & 0 & 0 & 0 & 0 & 0 & 0 & 0 & 0 \\ 
 $F_1^{(8)}$ & -0.0153 & 0.0274 & 0.1007 & -0.0020 & -0.1529 & -0.0432 & -0.0143 & 0.0321 \\
 $F_2^{(8)}$ & -0.0141 & 0.0567 & -0.1151 & -0.0485 & 0.1676 & -0.0472 & -0.0997 & 0.1429 \\
 $F_3^{(8)}$ & -0.0148 & 0.0414 & 0 & -0.0258 & 0 & 0 & -0.0118 & 0 \\
 $F_4^{(8)}$ & -0.1721 & 0.2192 & 0 & 0.1828 & 0 & 0 & 0.1006 & 0 \\
\end{tabular}
\end{ruledtabular}
\begin{ruledtabular}
\begin{tabular}{l|c  }
 $F$ & stability-matrix eigenvalues   \\ \hline
 $F_0^{(8)}$ & \{-2, \ -1, \ -1, \ 1, \ 1, \ 1, \ 0, \ 0\} \\
 $F_1^{(8)}$ & \hspace{2cm} \{4.1374, \ 2.1731, \ -2.0064, \ 1.3755, \ -1.1678, \ -0.9261, \ 0.2814, \ 0.1298\} \hspace{2cm}  \\
 $F_2^{(8)}$ &  \{4.2123, \ 2.3531, \ -2.0072, \ 1.5805, \ -1.1886, \ -0.9334, \ 0.2274, \ 0.1216\} \\
 $F_3^{(8)}$ &  \{3.6611, \ 2.8933, \ -2.0029, \ 1.685, \ -1.1024, \ -1.0081, \ 0.1351, \ -0.0505\} \\
 $F_4^{(8)}$ &  \{34.5986, \ 26.9475, \ 12.7525, \ 9.3877, \ 5.1825, \ 1.3058, \ -1.1215, \ -0.65\}
\end{tabular}
\end{ruledtabular}
\end{table}

\section{Stability analysis}
\label{staban}

In order to determine the stability properties of the fixed points one can analyze the flow
in their neighborhood where it is governed by the linearized system. 
For this purpose one calculates the eigenvalues of the 
stability matrix
\begin{gather}
\label{stm}
 (S_{ij}) \equiv \left( \frac{\partial \beta_i}{\partial \bar{p}_j} 
\right) \Bigl\vert_{\bar{p}=\bar{p}*} \ ,
\end{gather}
\noindent
where we denote the $n$ rescaled parameters of the Lagrangian by
$\bar{p} = \{\bar{p}_i \}$, a fixed point by $\{ \bar{p}_i^* \}$, and
the beta functions are given by 
$\beta_i(\bar{p}) \equiv k \partial_k \bar{p}_i$.
In general one obtains $n_s$ eigenvalues with positive real part,
$n_u$ with negative real part, and $n_m$ with vanishing real part.
The corresponding eigenvectors give rise to invariant subspaces of the
parameter space inside which the flow stays if one starts within them \cite{regev2006chaos}. 
In case of distinct eigenvalues there is a $n_s$-dimensional 
invariant subspace (called critical manifold) inside which the flow
is attracted towards the fixed point in the infrared limit $k=0$. 
Respectively, there exists a $n_u$-dimensional invariant subspace (called
unstable manifold) inside which the flow is repelled, and a $n_m$-dimensional
invariant subspace (called marginal manifold) inside which the flow has no
direction at all.
Here we note that complex valued eigenvalues always appear as conjugate pairs.
Referring to the real and imaginary parts of the associated complex eigenvectors as
eigenvectors, too, the critical manifold is spanned by $n_s$ eigenvectors, the
unstable manifold is spanned by $n_u$ eigenvectors, and the marginal manifold is
spanned by $n_m$ eigenvectors. 
Therefore, if $n_m=0$, one can reach the critical
manifold by tuning $n_u$ parameters starting anywhere in parameter space. 
Hence, a second-order phase transition with 
respect to a single scaling variable (temperature) can only exist
if we have exactly $n_u=1$. In this case we speak of an IR-stable fixed point.  \\
The stability-matrix eigenvalues are listed for each fixed point in
Table \ref{tabfp6}--\ref{tabfp8}.
We begin with discussing the LPA in sextic truncation order.
$F_1^{(6)}$ and $F_2^{(6)}$ are different, IR-stable, 
$U(2)_A \times U(2)_V$-symmetric spiral fixed points. $F_1^{(6)}$ is associated
with physical mass-matrix eigenvalues whereas $F_2^{(6)}$ is not. 
Their existence is highly nontrivial since they do not exist at quartic
truncation order, neither in the LPA \cite{Fukushima:2010ji,Fejos:2014qga}, nor in the LPA' \cite{Berges:2000ew}.
The critical exponent, $\nu \sim 1/1.3614 \sim 0.7345$, associated with
$F_1'^{(6)}$ is in unexpectedly good agreement with the values reported in Ref.\ \cite{Pelissetto:2013hqa}
($\nu \sim 0.71$ for the MZM scheme, $\nu \sim 0.76$ for the $\overline{MS}$ scheme). This 
agreement is most likely accidental. The value for the anomalous dimension is actually significantly
smaller ($\eta \sim 0.0334$ compared to $\eta \sim 0.1$). 
$F_3^{(6)}$ is an unstable $O(8)$-symmetric fixed point. All fixed points 
are also present in the LPA' without qualitative changes ($F_i^{(6)}$ corresponds
to $F_i'^{(6)}$). \\
Of particular interest to us are the marginal eigenvalues encountered for the
Gaussian fixed points, $F_0$, which will be discussed next. 
From a merely mathematical 
standpoint one can decide whether marginal eigenvalues are relevant
or not by going beyond the linear order utilized in Eq.\ (\ref{stm}). For this purpose
one can either use the second derivatives of the beta functions, or one has to perform 
a more general Lyapunov analysis. However, this is not meaningful in our case 
because in presence of
marginal eigenvalues one has to consider a change in the 
fixed-point structure at higher polynomial truncation order.
In general, such a change cannot be excluded by a nonlinear 
stability analysis at lower order.
The occurrence of the marginal eigenvalues, however, can be explained as follows.
In general the beta functions for a rescaled mass parameter $\bar{m}^2$, 
a rescaled quartic coupling 
$\bar{g}_4$, and a rescaled sextic coupling $\bar{g}_6$, respectively, are given by
\begin{gather}
\beta_{m^2} = (-2 + \eta) \bar{m}^2 + f_{2}(\bar{p}) \; , \;
\beta_4 = (D-4+2 \eta) \bar{g}_4 + f_{4}(\bar{p}) \; , \;
\beta_6 = (2D-6+3 \eta) \bar{g}_6 + f_{6} (\bar{p}) \; ,
\end{gather}
\noindent
where the $f_i (\bar{p})$ denote nonlinear functions of the rescaled parameters.
In FRG the polynomial order of these functions depends on the truncation order 
of the effective action, whereas in RG approaches based on a loop expansion
it depends on the loop order. Since these functions as well as the anomalous
dimension, $\eta$, vanish at the Gaussian fixed point, we can conclude that (for $D=3$) 
$\bar{m}^2$ and $\bar{g}_4$ are relevant parameters with respect to this fixed point.
They yield stability matrix eigenvalues $-2$ and $-1$, respectively. 
Similarly, the sextic coupling contributes a vanishing eigenvalue at the
Gaussian fixed point, and higher order couplings yield positive eigenvalues.
We conclude that the marginal eigenvalues in 
Table \ref{tabfp6}--\ref{tabfp8} do not render the stability 
analysis inconclusive. However, in the remainder of this section, we will
argue why the LPA' remains inconclusive, pointing out general differences between
FRG and other RG approaches first.
For a more fundamental comparison between both approaches we refer 
to Refs.\ \cite{Litim:2002xm,Codello:2013bra}. \\ 
In the framework of the $\epsilon$-expansion or other loop expansions
at fixed spatial dimension $D$, one usually argues 
that also in case
of non-Gaussian fixed points the canonical scaling dimension determines 
if a coupling can affect stability \cite{herbut2007modern}. 
Accordingly, depending on the sign of their 
canonical scaling dimension, one speaks of relevant, marginal, and irrelevant 
parameters. Obviously, especially marginal eigenvalues are sensitive to the 
loop order. Therefore, one has to consider the possibility that higher-order loop corrections
change the marginal eigenvalue into a nonvanishing one. 
It is important to note that if a marginal eigenvalue 
for a certain fixed point turns nonzero
at higher order, this can also change the stability properties of 
the other fixed points. This is for example the case in the $O(N=4)$ model
with di-icosahedral anisotropy. The $\epsilon$-expansion
of this model has been derived in Ref.\ \cite{Toledano:1985},
pointing out that the case of $N=4$ is special. In the presence of
an anisotropy, the $O(4)$-symmetric fixed point acquires a marginal eigenvalue
at one-loop order in the $\epsilon$-expansion whereas the anisotropic fixed 
point is IR-unstable. At two-loop order, however, the anisotropic fixed point
can become the IR-stable one. We reinvestigated the situation using
the FRG in LPA and 
found that the anisotropic fixed point also becomes IR-stable when going
beyond the quartic truncation order \cite{GrahlDiss}. \\
However, a change of stability can
occur even in the absence of any marginal eigenvalues. A famous example is the
$O(N)$ model with cubic anisotropy for $D=3$ \cite{Varnashev:1999ze,Pelissetto:2000ek}. The model exhibits 
an $O(N)$-symmetric (isotropic) fixed point as well as a cubic fixed point.
For $N > N_c$ the cubic fixed point is the IR-stable one, the isotropic
fixed point being IR-unstable, and vice versa for $N < N_c$.
The value for $N_c$ depends on the loop order as well as on the resummation scheme and
is still under debate. \\
In comparison to loop expansions, the stability matrix eigenvalues 
are much more sensitive to the polynomial truncation
order in the FRG formalism. 
Using FRG, the accuracy of the critical exponents heavily depends
on irrelevant couplings \cite{Litim:2002cf}.
This is explained by
the fact that fluctuations are taken into account differently in both approaches.
Irrelevant couplings can be safely ignored in the loop expansion and
nonperturbative effects are captured by using resummation.
In contrast, if we were able to solve the FRG equation without 
truncating the effective action, we would obtain exact results. 
In the LPA
at quartic truncation order, however, 
one generically reproduces the one-loop epsilon-expansion
results when setting the mass parameter to zero \cite{Kopietz:FRG,Fukushima:2010ji}. \\ 
Our conclusions are as follows.
Naively, one would trust the utilized approximation scheme
since no marginal eigenvalues appear for the non-Gaussian fixed points. However, we argued that
even in this case the fixed-point structure can change at higher truncation order. 
Especially the
presence of the unphysical fixed point advises caution. In fact, the spiral fixed points become
unstable fixed points ($F_1^{(8)}$ and $F_2^{(8)}$, respectively) at 
octic truncation order (see Table \ref{tabfp8}). Interestingly,
at this order one finds two unstable $O(8)$-symmetric fixed points. 
Going to any higher (finite) polynomial order in the LPA' will not clarify the situation.
If an IR-stable fixed point were found at higher order, one could not rule out its
disappearance beyond that order. And in the opposite case the 
discrepancy with Ref.\ \cite{Pelissetto:2013hqa} would require to go beyond the LPA' as well.
Therefore it is necessary to include derivative couplings in order to decide whether the 
$U(2)_A \times U(2)_V$-symmetric fixed point is stable or not. In addition, novel criteria to
assess the conclusiveness of truncation schemes need to be developed.

\section{Conclusions}
\label{conclusions}

We further investigated the possibility that the two-flavor chiral phase
transition can be of second order in the absence of the 
axial anomaly, using the FRG method 
in the LPA as well as in the LPA'. \\
We found two IR-stable, $U(2)_A \times U(2)_V$-symmetric fixed points
at sextic polynomial truncation order, one of them
associated with unphysical masses.
The value for the critical exponent, $\nu \sim 0.7345$, calculated 
for the one associated with physical masses is in (most likely accidental)
agreement with the result reported in Ref.\ \cite{Pelissetto:2013hqa}.
At higher polynomial order both fixed
points become unstable.
Nevertheless, the results of our research provide further evidence for the existence 
of the IR-stable, $U(2)_A \times U(2)_V$-symmetric fixed point from
an independent perspective. \\
The fact that an $U(2)_A \times U(2)_V$-symmetric
fixed point appears by simply including
sextic invariants demonstrates that its existence not only
depends on the spatial dimension but also on the 
way nonperturbative corrections are taken into account. In the 
framework of a resummed perturbative expansion this concerns the
resummation scheme and the perturbative order. \\
Our main conclusion is that the LPA' is not capable to unambiguously 
clarify the stability of the fixed points. 
Since the fixed-point structure of the dimensionally
reduced theory controls the behavior near $T_c$, previous 
finite-temperature studies \cite{Berges:2000ew,GrahlDiss,Fejos:2014qga} remain inconclusive, too.
We expect clarification beyond the LPA' taking into account derivative 
couplings. \\
Finally, the simultaneous occurrence of two IR-stable fixed points 
(although one of them being unphysical, and the truncation is 
not reliable) is interesting
regarding the universality hypothesis. The example illustrates that, in principle, it is possible that 
two systems sharing (a) the same spatial dimension, (b) the same number of order parameter components,
and (c) the same symmetry properties can be attracted to different IR-stable fixed points (here $F_1'^{(6)}$
and $F_2'^{(6)}$, respectively). Both associated universality classes are characterized by the same
representation of the same symmetry group. However, we state clearly that the given
example has to be regarded as an artifact of the utilized truncation. A similar situation, although to our
knowledge not strictly ruled out, is commonly not believed to appear in a physical setting.

\section*{Acknowledgment}

The author would like to thank HIC for FAIR for funding. The author would further like
to thank J{\"urgen} Eser, Francesco Giacosa, Mario Mitter, Dirk-Hermann Rischke, and
Bernd-Jochen Schaefer for valuable discussions.

%

\bibliographystyle{unsrt}
\bibliography{mybib_RG}

\begin{thebibliography}{10}

\bibitem{Leupold:2011zz}
S.~Leupold, K.~Redlich, M.~Stephanov, A.~Andronic, D.~Blaschke, et~al.
\newblock {Bulk properties of strongly interacting matter}.
\newblock {\em Lect.Notes Phys.}, 814:39--334, 2011.

\bibitem{Pawlowski:2010ht}
Jan~M. Pawlowski.
\newblock {The QCD phase diagram: Results and challenges}.
\newblock {\em AIP Conf.Proc.}, 1343:75--80, 2011.

\bibitem{Herbst:2013ufa}
Tina~Katharina Herbst, Mario Mitter, Jan~M. Pawlowski, Bernd-Jochen Schaefer,
  and Rainer Stiele.
\newblock {Thermodynamics of QCD at vanishing density}.
\newblock {\em Phys.Lett.}, B731:248--256, 2014.

\bibitem{Cossu:2013uua}
Guido Cossu, Sinya Aoki, Hidenori Fukaya, Shoji Hashimoto, Takashi Kaneko,
  et~al.
\newblock {Finite temperature study of the axial U(1) symmetry on the lattice
  with overlap fermion formulation}.
\newblock {\em Phys.Rev.}, D87(11):114514, 2013.

\bibitem{Bhattacharya:2014ara}
Tanmoy Bhattacharya, Michael~I. Buchoff, Norman~H. Christ, H.~T. Ding, Rajan
  Gupta, et~al.
\newblock {The QCD phase transition with physical-mass, chiral quarks}.
\newblock {\em Phys.Rev.Lett.}, 113:082001, 2014.

\bibitem{Pisarski:1983ms}
Robert~D. Pisarski and Frank Wilczek.
\newblock {Remarks on the Chiral Phase Transition in Chromodynamics}.
\newblock {\em Phys. Rev.}, D29:338--341, 1984.

\bibitem{Berges:1997eu}
J.~Berges, D.~U. Jungnickel, and C.~Wetterich.
\newblock {Two flavor chiral phase transition from nonperturbative flow
  equations}.
\newblock {\em Phys. Rev.}, D59:034010, 1999.

\bibitem{Berges:1998sd}
Juergen Berges, Dirk-Uwe Jungnickel, and Christof Wetterich.
\newblock {The chiral phase transition at high baryon density from
  nonperturbative flow equations}.
\newblock {\em Eur. Phys. J.}, C13:323--329, 2000.

\bibitem{Butti:2003nu}
Agostino Butti, Andrea Pelissetto, and Ettore Vicari.
\newblock {On the nature of the finite-temperature transition in QCD}.
\newblock {\em JHEP}, 08:029, 2003.

\bibitem{Calabrese:2004uk}
Pasquale Calabrese and Pietro Parruccini.
\newblock {Five loop epsilon expansion for U(n) x U(m) models: Finite
  temperature phase transition in light QCD}.
\newblock {\em JHEP}, 0405:018, 2004.

\bibitem{Braun:2010vd}
Jens Braun, Bertram Klein, and Piotr Piasecki.
\newblock {On the scaling behavior of the chiral phase transition in QCD in
  finite and infinite volume}.
\newblock {\em Eur.Phys.J.}, C71:1576, 2011.

\bibitem{Fukushima:2010ji}
Kenji Fukushima, Kazuhiko Kamikado, and Bertram Klein.
\newblock {Second-order and Fluctuation-induced First-order Phase Transitions
  with Functional Renormalization Group Equations}.
\newblock {\em Phys.Rev.}, D83:116005, 2011.

\bibitem{Grahl:2013pba}
Mara Grahl and Dirk~H. Rischke.
\newblock {Functional renormalization group study of the two-flavor linear
  sigma model in the presence of the axial anomaly}.
\newblock {\em Phys.Rev.}, D88:056014, 2013.

\bibitem{Pelissetto:2013hqa}
Andrea Pelissetto and Ettore Vicari.
\newblock {Relevance of the axial anomaly at the finite-temperature chiral
  transition in QCD}.
\newblock {\em Phys.Rev.}, D88:105018, 2013.

\bibitem{Aoki:2013zfa}
Sinya Aoki, Hidenori Fukaya, and Yusuke Taniguchi.
\newblock {1st or 2nd; the order of finite temperature phase transition of
  $Nf=2$ QCD from effective theory analysis}.
\newblock {\em PoS}, LATTICE2013:139, 2013.

\bibitem{Meggiolaro:2013swa}
Enrico Meggiolaro and Alessandro Morda.
\newblock {Remarks on the U(1) axial symmetry and the chiral transition in QCD
  at finite temperature}.
\newblock {\em Phys.Rev.}, D88:096010, 2013.

\bibitem{Nakayama:2014sba}
Yu~Nakayama and Tomoki Ohtsuki.
\newblock {Bootstrapping phase transitions in QCD and frustrated spin systems}.
\newblock 2014.
\newblock arXiv:1407.6195.

\bibitem{Fejos:2014qga}
G.~Fejos.
\newblock {Fluctuation induced first order phase transition in U(n)xU(n) models
  using chiral invariant expansion of FRG flows}.
\newblock 2014.
\newblock arXiv:1409.3695.

\bibitem{Paterson:1980fc}
A.~J. Paterson.
\newblock {Coleman-Weinberg Symmetry Breaking In The Chiral SU(n) x SU(n)
  Linear $\sigma$ Model}.
\newblock {\em Nucl. Phys.}, B190:188, 1981.

\bibitem{GrahlDiss}
Mara Grahl.
\newblock {\em Low-energy effective models for two-flavor quantum
  chromodynamics and the universality hypothesis}.
\newblock PhD thesis. Univ.-Bibliothek Frankfurt am Main,
  urn:nbn:de:hebis:30:3-337944, 2014.

\bibitem{Patkos:2012ex}
A.~Patkos.
\newblock {Invariant formulation of the Functional Renormalisation Group method
  for $U(n)\times U(n)$ symmetric matrix models}.
\newblock {\em Mod.Phys.Lett.}, A27:1250212, 2012.

\bibitem{Pawlowski:1996ch}
J.M. Pawlowski.
\newblock {Exact flow equations and the U(1) problem}.
\newblock {\em Phys.Rev.}, D58:045011, 1998.

\bibitem{Fischer:2011pk}
Christian~S. Fischer and Jens~A. Mueller.
\newblock {On critical scaling at the QCD $N_f=2$ chiral phase transition}.
\newblock {\em Phys.Rev.}, D84:054013, 2011.

\bibitem{Schaefer:2013isa}
Bernd-Jochen Schaefer and Mario Mitter.
\newblock {Three-flavor chiral phase transition and axial symmetry breaking
  with the functional renormalization group}.
\newblock {\em Acta Phys.Polon.Supp.}, 7(1):81--90, 2014.

\bibitem{Mitter:2013fxa}
Mario Mitter and Bernd-Jochen Schaefer.
\newblock {Fluctuations and the axial anomaly with three quark flavors}.
\newblock {\em Phys.Rev.}, D89:054027, 2014.

\bibitem{Aoki:2012yj}
Sinya Aoki, Hidenori Fukaya, and Yusuke Taniguchi.
\newblock {Chiral symmetry restoration, eigenvalue density of Dirac operator
  and axial U(1) anomaly at finite temperature}.
\newblock {\em Phys.Rev.}, D86:114512, 2012.

\bibitem{Sharma:2013nva}
Sayantan Sharma, Viktor Dick, Frithjof Karsch, Edwin Laermann, and Swagato
  Mukherjee.
\newblock {Investigation of the $U_A(1)$ in high temperature QCD on the
  lattice}.
\newblock 2013.
\newblock arXiv:1311.3943.

\bibitem{Jungnickel:1995fp}
D.~U. Jungnickel and C.~Wetterich.
\newblock {Effective action for the chiral quark-meson model}.
\newblock {\em Phys. Rev.}, D53:5142--5175, 1996.

\bibitem{Przy}
J.~{Przystawa}.
\newblock {Symmetry and phase transitions}.
\newblock {\em Physica A Statistical Mechanics and its Applications},
  114:557--563, August 1982.

\bibitem{DimRed2ndorder}
S.~Bornholdt, N.~Tetradis, and C.~Wetterich.
\newblock {High temperature phase transition in two scalar theories}.
\newblock {\em Phys.Rev.}, D53:4552--4569, 1996.

\bibitem{Kopietz:FRG}
Peter Kopietz, Lorenz Bartosch, and Florian Sch{\"u}tz.
\newblock {\em Introduction to the functional renormalization group}.
\newblock Lecture Notes in Physics. Springer, Berlin, 2010.

\bibitem{Berges:2000ew}
Juergen Berges, Nikolaos Tetradis, and Christof Wetterich.
\newblock {Nonperturbative renormalization flow in quantum field theory and
  statistical physics}.
\newblock {\em Phys.Rept.}, 363:223--386, 2002.

\bibitem{regev2006chaos}
O.~Regev.
\newblock {\em Chaos and Complexity in Astrophysics}.
\newblock Cambridge University Press, 2006.

\bibitem{Litim:2002xm}
Daniel~F. Litim and Jan~M. Pawlowski.
\newblock {Completeness and consistency of renormalisation group flows}.
\newblock {\em Phys.Rev.}, D66:025030, 2002.

\bibitem{Codello:2013bra}
Alessandro Codello, Maximilian Demmel, and Omar Zanusso.
\newblock {Scheme dependence and universality in the functional renormalization
  group}.
\newblock {\em Phys.Rev.}, D90:027701, 2014.

\bibitem{herbut2007modern}
I.~Herbut.
\newblock {\em A Modern Approach to Critical Phenomena}.
\newblock Cambridge University Press, 2007.

\bibitem{Toledano:1985}
J.-C. Toledano~et al.
\newblock {Renormalization-group study of the fixed points and of their
  stability for phase transitions with four-component order parameters}.
\newblock {\em Phys.Rev.}, B31:7171, 1985.

\bibitem{Varnashev:1999ze}
K.B. Varnashev.
\newblock {Stability of a cubic fixed point in three-dimensions: Critical
  exponents for generic N}.
\newblock {\em Phys.Rev.}, B61:14660, 2000.

\bibitem{Pelissetto:2000ek}
Andrea Pelissetto and Ettore Vicari.
\newblock {Critical phenomena and renormalization group theory}.
\newblock {\em Phys.Rept.}, 368:549--727, 2002.

\bibitem{Litim:2002cf}
Daniel~F. Litim.
\newblock {Critical exponents from optimized renormalization group flows}.
\newblock {\em Nucl.Phys.}, B631:128--158, 2002.

\end{thebibliography}


\end{document}